# HIGH-ENERGY PARTICLE COLLIDERS: PAST 20 YEARS, NEXT 20 YEARS, AND BEYOND*

V.Shiltsev#, Fermilab, Batavia, IL 60510, USA

*Abstract*

Particle colliders for high-energy physics have been in the forefront of scientific discoveries for more than half a century. The accelerator technology of the colliders has progressed immensely, while the beam energy, luminosity, facility size, and cost have grown by several orders of magnitude. The method of colliding beams has not fully exhausted its potential but has slowed down considerably in its progress. This paper briefly reviews the colliding beam method and the history of colliders, discusses the development of the method over the last two decades in detail, and examines near-term collider projects that are currently under development. The paper concludes with an attempt to look beyond the current horizon and to find what paradigm changes are necessary for breakthroughs in the field.

## PAST AND PRESENT COLLIDERS

It is estimated that in the post-1938 era, accelerator science has influenced almost 1/3 of physicists and physics studies and on average contributed to physics Nobel Prize-winning research every 2.9 years [1]. Colliding beam facilities which produce high-energy collisions (interactions) between particles of approximately oppositely directed beams did pave the way for progress since the 1960's. Discussion in this section mainly follows recent publication [2].

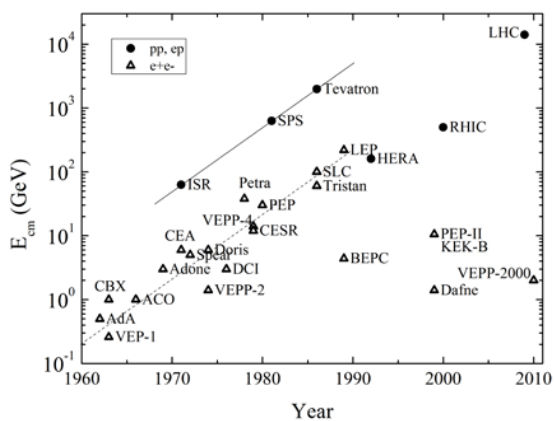

Figure 1: Colliders over the decades (after [2]).

Twenty nine colliders reached operational stage between the late 50's and now. The energy of colliders has been increasing over the years as demonstrated in Fig.1. There, the triangles represent maximum CM energy and the start of operation for lepton (usually, $e+e-$) colliders and full circles are for hadron (protons, antiprotons, ions, proton-electron) colliders. One can see that until the early 1990's, the CM energy on average increased by a factor of 10 every decade and, notably, the hadron colliders were 10-20 times more powerful. Since then, following the demands of high energy physics, the paths of the colliders diverged to reach record high energies in the particle reaction. The Large Hadron Colider (LHC) was built at CERN, while new $e+e-$ colliders called "particle factories" were focused on detail exploration of phenomena at much lower energies. Over the last 50 years, the luminosity of the colliders has improved by more than 6 orders of magnitude and reached record high values of over $10^{34}$cm$^{-2}$s$^{-1}$. At such luminosity, one can expect to produce, e.g., 100 events over one year of operation (about $10^7$ s) if the reaction cross section is 1 picobarn (pb)=$10^{-39}$ cm$^2$.

**Table 1**: Past, present and possible future colliders; hadron colliders are in **bold**, lepton colliders in *Italic*, facilities under construction or in decisive design and planning stage are listed in parenthesis (…)

|  | early 1990's | early 2010's | 2030's |
|---|---|---|---|
| Europe | H**E**RA, (**LHC**) *LEP* (*Dafne*) | **LHC** (*Super-B*, **HL-LHC, LH**e**C, E**NC**) | **HE-LHE** *CLIC?* |
| Russia | *VEPP2, VEPP4* (**UNK**, *VLEPP*) | *VEPP2000, VEPP-4M* (**NICA**, *Tau-C*) | **NICA** ? *Higgs Factory* ? |
| US | **Tevatron**, (**SSC**) *SLC, CESR*,(*PEPII*) | **RHIC** (*e***RHIC**, *ELIC*) | *Muon Collider* ? *PWLA/DLA* ? |
| Asia | *Tristan, BEPC* (*KEK-B*) | *BEPC* (*Super-KEKB*) | *ILC* ? *Higgs Factory* ? |
| Total | 9 (7) | 5 (9) | 1 + ? |

At the same time, the number of the facilities in operation has dropped from 9 to 5, as indicated in Table 1 which lists all operational colliders as of the early 1990's and now (early 2010's) and accounts for the projects under construction or under serious consideration at this time (in parenthesis).

## NEXT TWO DECADES

There are several colliders currently under construction or at the design stage and which, therefore, have good prospects of becoming operational and deliver results in the next 20 years. Schematically they can be categorized by as follows:

*Energy Frontier :* the LHC luminosity upgrade project HL-LHC [3] to quintuple the performance of the energy frontier machine by mid-2020's to $5 \cdot 10^{34}$cm$^{-2}$s$^{-1}$ with luminosity leveling at 14 TeV c.m. energy in proton-proton collisions; the LHC energy upgrade to ~33 TeV cme will require development 20T dipoles [4].

*Low-energy hadron collders:* like, e.g., Nuclotron-based Ion Collider fAcility (NICA) currently under construction at JINR (Dubna, Russia) [5] which will allow for the study of ion-ion ($Au^{+79}$) and ion-proton collisions in the energy range of 1-4.5 GeV/amu with average luminosity of $10^{27}$ cm$^{-2}$ s$^{-1}$ and also polarized proton-proton (5-12.6 GeV) and deuteron-deuteron (2-5.8 GeV/amu) collisions.

*Electron-hadron collders:* such as LHeC [6], in which polarized electrons of 60 GeV to possibly 140 GeV collide with LHC protons of 7000 GeV with design luminosity of about $10^{33}$ cm$^{-2}$s$^{-1}$; eRHIC at BNL (polarized nuclear beam to 100 GeV/nucleon and polarized protons upto 250 GeV colliding with 20-30 GeV ERL-accelerated polarized electrons, the luminosity varies from $10^{33}$ cm$^{-2}$s$^{-1}$ to $10^{34}$ cm$^{-2}$s$^{-1}$) [7] and Electron-Ion Collider (ELIC) at JLab (30-7 GeV polarized electrons and 30 to 150 GeV storage ring for ions, high luminosities of ~$5 \cdot 10^{33}$ cm$^{-2}$s$^{-1}$ to $10^{35}$ cm$^{-2}$s$^{-1}$ require continuous electron cooling) [8]; and electron-nucleon collider ENC at FAIR (GSI, Darmstadt, Germany) utilizing the 15 GeV antiproton high-energy storage ring HESR for polarized $p$ and $d$ beams, with the addition of a polarized $e$- 3.3 GeV storage ring with peak luminosities in the range of $10^{32}$ to $10^{33}$cm$^{-2}$s$^{-1}$ [9].

*Electron-positron factories:* such as asymmetric $e+e-$ Super-$B$ factory in KEK [10] with beam energies of about 4 GeV and 7 GeV and with a design luminosity approaching $10^{36}$ cm$^{-2}$ s$^{-1}$; and Tau-Charm factory in Novosibirsk[11] which calls for c.m. collision energy variable from 3 GeV to 4.5 GeV (from *J/psi* resonances to charm barions) and luminosity in excess of $10^{35}$ cm$^{-2}$ s$^{-1}$.

*Higgs Factories:* $e+e-$ colliders with an optimal $E_{cm}$ ~ $m_H$ + (110±10) GeV ~250 GeV can be either ring-based as, e.g., TLEP [12] or be based on the ILC-type linear collider [13]; their issues and challenges are discussed in [14]. A possible alternative can be a $\mu+\mu-$ collider with factor of two lower c.m. energy $E_{cm}$~$m_H$ which would have advantageously large cross section some $(m_\mu/m_e)^2$~40,000 times higher than for $e+e-$, and significantly smaller c.m.e. spread $\delta E_{cm}/E_{cm}$~ 0.01-0.003% (compared to ~0.2% for the $e+e-$ factories) [15].

*Energy Frontier Lepton Colliders:* if - as presently believed - a multi-TeV lepton collider will be needed to follow the LHC discoveries, then the most viable options currently under consideration are $e+e-$ linear colliders ILC and CLIC [16] or $\mu+\mu-$ Muon Collider [15] – see Table 2. Each of these options has its own advantages and issues [17].

## FAR FUTURE COLLIDERS

The forecast on the far future of colliders beyond, say, the 2030's, will require several things: understanding of the available resources, the desired science reach, and on the possibilities. As of today, the world's particle physics research budget is some 3B$. Assuming that it will not change by much in the future (in todays' dollars) and that not more than 1/3 of the budget can be dedicated to construction of the next energy frontier collider, one can estimate the cost of an affordable future facility to be about or less then 10B$ (in current prices). Other desired features of such flagship could be: it ideally should not exceed 10 or few 10's of km in length, it should stay under $O$(100MW) of AC wall power consumption, and, of course, it would be great if its energy reach significantly (> ×10) exceeds that of the LHC.

**Table 2**: Comparison of Lepton Collider alternatives

|  | ILC | CLIC | MC |
|---|---|---|---|
| c.m.e., TeV | 0.5 | 3 | 1.5-4 |
| $dE/E$, rms | ~2% | >5% | ~0.1% |
| $L$, cm$^{-2}$s$^{-1}$ | $2 \cdot 10^{34}$ | $2 \cdot 10^{34}$ * | $(1-4) \cdot 10^{34}$ |
| Feasibility | 2007 | 2012 | ~2016 |
| Techn.design | 2013 | 2016 | ~2020 |
| # elements | 38,000 | 260,000 | 10,000 |
| Σ length, km | 36 | ~60 | 14-20 |
| AC pwr, MW | 230 | 580 | ~170-220 |

* peak luminosity within 1% c.m. energy spread

The cost estimates of the modern colliders are quite complicated, but an attempt to analyse known costs of 15 large accelerators (SSC, RHIC, VLHC, ILC, B-Factories, FNAL MI, NLC, SNS, LHC, XFEL, FAIR, ESS, CLIC, SPL, NuFactory) has resulted in a phenomenological cost model [18] that matches the actual costs (in terms of *TPC*, the Total Project Cost, widely used in the US DOE accounting ) within approximately ±30% :

$$TPC = \alpha \left( \frac{L}{10km} \right)^{1/2} + \beta \left( \frac{E}{1TeV} \right)^{1/2} + \gamma \left( \frac{P}{100MW} \right)^{1/2}$$

where, $L$ is the total length of the tunnel, $E$ is cm energy, $P$ is total site AC power for the facility; and technology dependent coefficients are $\alpha$=2B$, $\beta$=1B$ for NC magnets, 3B$ for SC magnets, and ≈10B$ for NC/SC RF, $\gamma$=2B$ - see Fig.2.

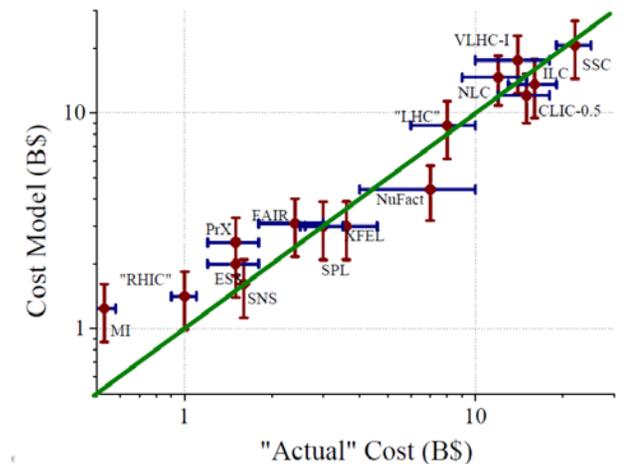

Figure 2: Phenomenological cost model vs actual cost for large accelerator facilities.

The model predicts that if currently available technologies are employed, then only Higgs factories or a few-TeV Muon Collider can potentially meet the desired cost requirement. If the costs of civil construction and power infrastructure are not change, i.e., $\alpha$ and $\gamma$ are as of today, then future 100 TeV, 10 km and 100MW collider should be based on technology that guarantees cost of the accelerator components of about 6B$, i.e., the coefficient $\beta <$ 0.6B$, that is two times cheaper than for modern NC magnets. Are there now or can we think of such technologies?

## ARE THERE WAYS TO THE FUTURE ?

To the date, the community thinking was more about fast acceleration to get to the energies of interest within a smaller footprint, rather than explicitly about the cost. The opportunities for high energy colliders under active discussion now are :

*Acceleration in dielectric structures:* direct beam excitation of wakefields in dielectric structures allows accelerating gradients of ~100 MV/m and ~1 GV/m at the microwave $O$(10) GHz and THz ranges [19]. Conceptually, a DWA-based linear collider would consist of a large number of ~100 GeV modules (stages) with some ~0.3 GeV/m gradient each driven by a separate ~1 GeV high intensity electron beam. Even without going into difficulties associated with staging, cost and power considerations, it is hard to imagine that more than 3 TeV c.m. energy DWA facility can fit within in a 10 km site. Further increase of the gradient to ~1-3 GV/m is thought to be possible in μm scale dielectric structures driven by lasers operating in optical or near-infrared regime [20].

*Acceleration in plasma:* Generation of the PWA gradients on the order of 30-100 GV/m by beams or lasers in plasma densities of $n_0$=$10^{17}$-$10^{18}$ cm$^{-3}$ have been already demonstrated in small scale (few cm to a meter) experiments [21, 22]. Laser-PWA collider would operate at relatively low densities $n_0$~$10^{17}$cm$^{-3}$ and energy gain per ~2m stage of 10 GeV and average gradient ~5TeV/km, while a beam-PWA collider design needs a conventional high power 25 GeV electron drive beam accelerator and many (dozens to hundreds, depending on the final energy) 1-m long plasma cells with an average geometric accelerating gradient ~0.25 TeV/km. Without going into the luminosity considerations [23] and just projecting such gradients further one can think of an ultimate ~10 TeV $e+e-$ collider within 10 km footprint.

More than an order of magnitude reduction of the cost of accelerator components in such colliders is predicted to be needed in order to fit under the expected TPC limit [18]. In addition, due to radiation in the focusing channel and beamstrahlung at IPs acceleration of electrons is limited to only few TeV. Heavier particles are required.

*Acceleration in crystal channels:* field gradients of upto 100 GeV/cm or 10 TV/m are possible in solids due to high density of charge carriers $n_0$~$10^{22-23}$ cm$^{-3}$. Muons, which do not interact with nuclei, would the particles of choice. X-ray lasers can efficiently excite solid plasma and accelerate muons inside a crystal channel waveguide [24], though ultimate acceleration gradients ~10TeV/m might require relativistic intensities, exceeding those conceivable for X-ray sources as of today. Side injection of powerful x-ray pulses into continuous fiber of 0.1 – 10 km long fiber allows to avoid multiple staging issues intrinsic to other methods and reach 10-1000 TeV collider energies – see Fig.3.

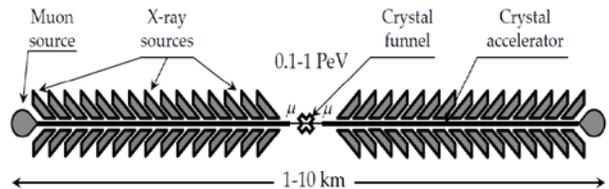

Figure 3: Layout of linear X-ray crystal muon collider [2].

It is not possible to even guess-timate the cost of such unproven acceleration method. One can only foresee that a quest for 100-1000 TeV energies will come at the price of the expected luminosities and will require at least three paradigm shifts [2]: 1) development of the new economical technology for ultrahigh acceleration gradients ~0.1-10 TeV/m in sub-nm structures (crystal fibers); 2) acceleration of heavier particles, preferably, muons; and 3) new approaches to HEP at such energies research with luminosities limited to ~$10^{30-32}$ cm$^{-2}$s$^{-1}$.

## REFERENCES

[1] E. Haussecker, A.W.Chao, *Phys.Persp.* **13** 146 (2011)
[2] V.Shiltsev, *Phys. Uspekhy* **55** 965–976 (2012)
[3] L.Rossi, in Proc. IPAC 2011 (Spain), p.908
[4] HE-LHC, Preprint CERN-2011-003 (2011)
[5] G.Trubnikov, et al., in Proc. 2010 RuPAC, p.14
[6] M.Klein, in Proc. IPAC 2011 (Spain), p.908
[7] V.Ptitsyn, et al. in Proc. IPAC 2011(Spain), p.3726
[8] S.Ahmed, et al., in Proc. PAC'11 (New York), p.2306
[9] A.Lehrach, et al., *J.Phys.:Conf.Ser.***295** 012156 (2011)
[10] T.Abe, et al., arxiv:1011.0352
[11] E.Levichev, *Phys.Part.Nucl. Lett.* **5** 554 (2008)
[12] Koratzinos, M. et al. arXiv:1305. 6498 (2013)
[13] *ILC TDR*, FERMILAB-TM-2554 (2013)
[14] S.Henderson, these Proceedings
[15] S.Geer, Annu.Rev.Nucl.Part.Sci. **59**, p.347 (2009)
[16] J.P.Delahaye, *Mod. Phys. Lett. A* **26** 2997 (2011)
[17] V.Shiltsev, *Mod. Phys. Lett. A* **25** 567 (2010)
[18] V.Shiltsev, in Proc. "Pre-Snowmass Workshop on Frontier Capabilities " (U.Chicago, Feb 25-26, 2013)
[19] W.Gai, *AIP Conf. Proc.* **1086** 3 (2008)
[20] G. Travish, et al, *AIP Conf. Proc.* **1086**, pp. 502–507.
[21] C.Schroeder, et al *PRSTAB* **13** 1013014 (2010)
[22] A.Seryi, *Nucl. Instr. Meth A* **623** 23 (2010)
[23] V.Lebedev, S.Nagaitsev, PRSTAB **16** (2013)
[24] T.Tajima, M.Cavenago *PRL* **59** 1440 (1987)